\documentclass[%
aip,
pof,
preprint,
]{revtex4-1}

\usepackage{graphicx}%
\usepackage{amssymb}%
\usepackage{xspace}%

\newcommand{\micron}{$\mu$m\xspace}     

\begin{document}

\title{Bubble size prediction in co-flowing streams}

\author{Wim van Hoeve} \affiliation{Physics of Fluids, Faculty of Science and Technology, and MESA$^+$ Institute for Nanotechnology, University of Twente, P.O.~Box 217, 7500 AE Enschede, The Netherlands}
\author{Benjamin Dollet} \affiliation{Institut de Physique de Rennes, UMR UR1-CNRS 6251, Universit\'e de Rennes 1, Campus de Beaulieu, B\^atiment 11A, F-35042 Rennes Cedex, France}
\author{Jos\'{e} M.~Gordillo} \affiliation{\'{A}rea de Mec\'{a}nica de Fluidos, Departamento de Ingenier\'{\i}a Aeroespacial y Mec\'{a}nica de Fluidos, Universidad de Sevilla, \mbox{Avda.\ de los Descubrimientos s/n}, 41092, Sevilla, Spain}
\author{Michel Versluis}
\author{Detlef Lohse} \affiliation{Physics of Fluids, Faculty of Science and Technology, and MESA$^+$ Institute for Nanotechnology, University of Twente, P.O.~Box 217, 7500 AE Enschede, The Netherlands}


\date{10 February 2011}

\begin{abstract}
In this paper, the size of bubbles formed through the breakup of a gaseous jet in a co-axial microfluidic device is derived. The gaseous jet surrounded by a co-flowing liquid stream breaks up into monodisperse microbubbles and the size of the bubbles is determined by the radius of the inner gas jet and the bubble formation frequency. We obtain the radius of the gas jet by solving the Navier-Stokes equations for low Reynolds number flows and by minimization of the dissipation energy. The prediction of the bubble size is based on the system's control parameters only, \emph{i.e.}~the inner gas flow rate $Q_\mathrm{i}$, the outer liquid flow rate $Q_\mathrm{o}$, and the tube radius $R$. For a very low gas-to-liquid flow rate ratio ($Q_\mathrm{i} / Q_\mathrm{o} \rightarrow 0$) the bubble radius scales as $r_\mathrm{b} / R \propto \sqrt{ Q_\mathrm{i} / Q_\mathrm{o} }$, independently of the inner to outer viscosity ratio $\eta_\mathrm{i}/\eta_\mathrm{o}$ and of the type of the velocity profile in the gas, which can be either flat or parabolic, depending on whether high-molecular-weight surfactants cover the gas-liquid interface or not. However, in the case in which the gas velocity profiles are parabolic and the viscosity ratio is sufficiently low, \emph{i.e.}~$\eta_\mathrm{i}/\eta_\mathrm{o}\ll 1$, the bubble diameter scales as  $r_\mathrm{b}\propto (Q_\mathrm{i}/Q_\mathrm{o})^\beta$, with $\beta$ smaller than $1/2$.
\end{abstract}

\maketitle

\section{Introduction}
The controlled formation of monodisperse microbubbles and microdroplets at high production rates is important in many industrial and medical applications. For example, the food industry seeks new methods to generate \emph{en masse} monodisperse droplets and bubbles to accurately control the density and structure of a wide variety of materials \cite{vanderGraaf2005,Hettiarachchi2010}. In medical ultrasound imaging, microbubbles can be used as ultrasound contrast agents (UCA), where the resonance size of the bubble determines its acoustic response \cite{Klibanov2002,Hettiarachchi2007}.
Commercially available UCA are produced using ultrasound-induced bubble formation methods, which results in a wide size distribution, \emph{i.e.}~UCA have a mean of 2\,\micron bubbles, but bubbles with a size between 1\,\micron and 20\,\micron exist in the population. Consequently, resonance occurs only for a small selection of bubbles. Improving the sensitivity in diagnostic imaging can thus be achieved by narrowing the size distribution so that more bubbles are at resonance size.

\begin{figure}
    \includegraphics[]{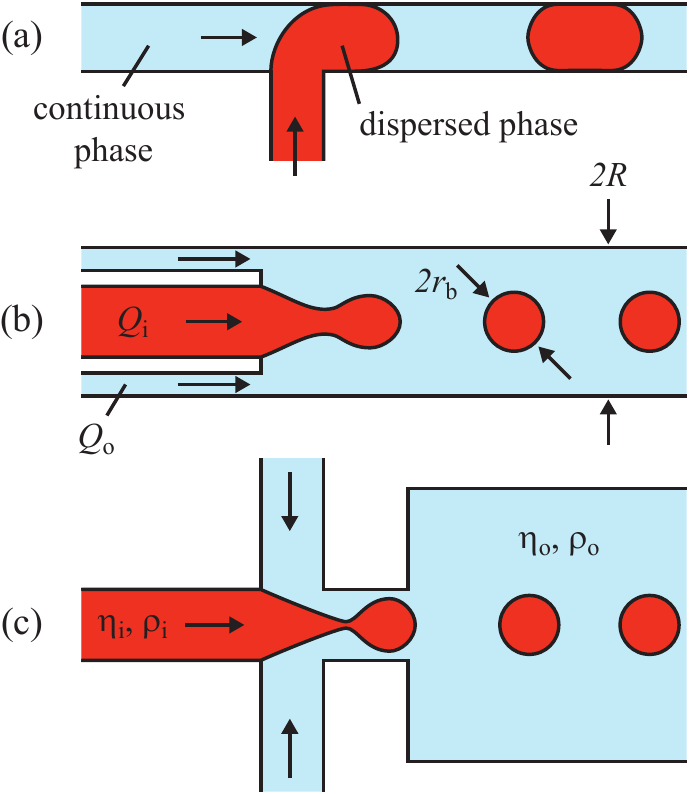}
    \caption{\label{Fig:CH4:microdevices}Schematic representation of the three main microfluidic geometries used for the formation of droplet and bubbles. The dispersed phase is injected (a) in a cross-flowing stream through a T-shaped junction, (b) in a co-flowing stream, and (c) in a focused stream imposed by the continuous phase.}
\end{figure}
The use of microfluidic technology for the generation of these accurately produced droplets and bubbles has received considerable attention recently \cite{Gunther2006,Christopher2007,Baroud2010,Nazir2010}.
Various microfluidic geometries are used, for example, T-shaped devices \cite{Thorsen2001,Chen2009}, co-flow, and flow-focusing geometries \cite{Ganan-Calvo1998,Ganan-Calvo2001,Anna2003}, as is shown schematically in fig.~\ref{Fig:CH4:microdevices}.
In the latter, the dispersed phase is focused by the outer continuous phase to enter a narrow channel (see fig.~\ref{Fig:CH4:microdevices}c).
Droplets (or bubbles) are formed in the `dripping' mode or `jetting' mode depending on the system's parameters. These include the inner and outer volumetric flow rates, $Q_\mathrm{i}$ and $Q_\mathrm{o}$, the channel dimensions, and the fluid parameters (viscosity and surface tension).
When the outer continuous flow rate $Q_\mathrm{o}$ is low, the droplets are generated in the dripping regime. In this regime, the dispersed phase enters the narrow channel and almost completely blocks the continuous phase. This leads to the formation of a neck, connecting the droplet to the inner feeding channel, that is gradually squeezed by the outer fluid until it breaks and a droplet is released \cite{Garstecki2005}. In Raven \emph{et al.}~\cite{Raven2006} this mode of operation was used to generate a wide variety of foams. For increasing gas fraction ($Q_\mathrm{i} / Q_\mathrm{o}$) foams are produced that consist of separated bubbles, to bubbly flow, to dry bamboo foam.
Droplet formation in the dripping regime is characterized by droplets with a size comparable to the size of the channel and its droplet production rate is typically low.

When the outer liquid velocity is sufficiently high, such that the capillary number of the outer fluid $\textrm{Ca}_\mathrm{o} = \eta_\mathrm{o} \tilde{u} / \gamma \ge 1$, with outer liquid viscosity $\eta_\mathrm{o}$, velocity at the interface $\tilde{u}$, and surface tension $\gamma$, viscosity overcomes surface tension forces and an elongated jet is formed \cite{Utada2007,Ganan-Calvo2007,Marin2008,Castro2009}. This regime is referred to as the jetting regime. The cylindrical jet breaks up in equally-sized fragments driven by the classical mechanism of droplet formation through a Rayleigh-Plateau instability driven by surface tension forces \cite{Guillot2007}. High-throughput monodisperse droplet formation in the jetting mode is of great value for industrial applications where a high-production rate is essential.

\begin{figure}[t]
    \includegraphics[]{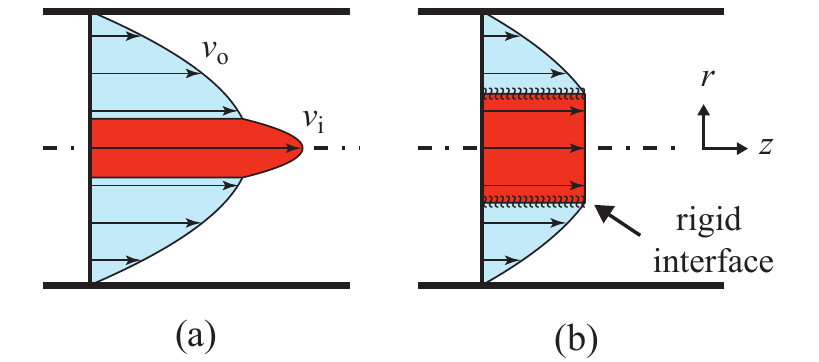}
    \caption{\label{Fig:CH4:velocityprofiles}Velocity profiles for the inner and outer flow in a co-flowing stream. (a) Poiseuille velocity profile for both inner and outer flow for two co-flowing liquids, with $\eta_\mathrm{i} < \eta_\mathrm{o}$. (b) The inner gas flow shows a flat velocity profile, whereas the outer flow shows a Poiseuille profile. The rigid interface allows for a discontinuity of shear stress at the gas-liquid surface. The red core represents the inner phase.}
\end{figure}

Utada \emph{et al.} \cite{Utada2005} and Utada \emph{et al.} \cite{Utada2007} presented a simple description of the droplet size in a liquid-liquid co-flowing microfluidic system with a geometry similar to that considered here. In their work, the inner liquid viscosity $\eta_\mathrm{i}$ is lower than the outer liquid viscosity $\eta_\mathrm{o}$ ($\eta_\mathrm{i}/\eta_\mathrm{o}\simeq 0.1$) and both the inner and outer flows are pressure driven; consequently, they exhibit Poiseuille-like velocity profiles (see fig.~\ref{Fig:CH4:velocityprofiles}a).

In this work we study microbubble formation through the breakup of a cylindrical jet in an axisymmetric gas-liquid co-flowing device formed by two coaxial tubes of circular section. Differently to the cases considered by Utada et al.~\cite{Utada2005,Utada2007}, we assume here an inner gas flow with negligible viscosity ($\eta_\mathrm{i} \ll \eta_\mathrm{o}$). Two different situations are considered depending upon whether the gas-liquid interface is free or rigidified by high-molecular-weight surfactants such as phospholipids, amphiphilic polymers, \emph{etc.} These complex molecules are highly relevant in UCA microbubble formation \cite{Hettiarachchi2010} to avoid gas dissolution into the carrier liquid.
Notice that the main difference between the two situations studied here is that high-molecular-weight surfactants naturally populate the gas-liquid interface and form a rigid interface that greatly affect the inner flow boundary conditions (see fig.~\ref{Fig:CH4:velocityprofiles}b).

In this paper we obtain a complete description of the radius of the inner gaseous jet and bubble size solely based on: the gas and liquid flow rates, the size of the channel, and the liquid properties.
We first describe droplet and bubble formation in the absence of high-molecular-weight surfactants, following the calculation already presented in Utada \emph{et al.}~\cite{Utada2005}.
We do so in preparation for the key new finding of this paper in which bubble formation with high-molecular-weight surfactants at the interface is described. We use minimization of energy dissipation to close the system of equations describing these specific conditions \cite{Stauffer2005}.

\section{Droplet and bubble formation from a liquid or gas jet\label{Sec:CH4:liquidliquidsystem}}
In order to be able to highlight the analogies and differences between gas-liquid and liquid-liquid systems, we first repeat and elucidate the essence of the droplet-size calculation of Utada \emph{et al.}~\cite{Utada2005} and then we extend this result to the case of bubbles.
Let us consider the flow of two immiscible fluids without surfactants in an axisymmetric co-flow device, as depicted in fig.~\ref{Fig:CH4:microdevices}b. These type of devices are typically fabricated by careful alignment of two coaxial capillaries: a tapered inner capillary for the supply of the dispersed phase and an outer capillary with radius $R$ that delivers the continuous phase. When the outer liquid velocity is sufficiently high, such that viscous forces overcome surface tension forces ($\textrm{Ca}_\mathrm{o} \ge 1$), an elongated liquid jet is formed. In the case of the dispersed phase is a liquid, the cylindrical jet ultimately develops undulations driven by surface tension that lead to the jet disruption into droplets with sizes comparable to the jet diameter. This breakup mechanism is known as Rayleigh breakup \cite{Rayleigh1879}. The size of the droplets is $V = \lambda^\star \pi {\tilde{r}}^2$, where $\lambda^\star$ is the wavelength of the fastest growing disturbance and ${\tilde{r}}$ the unperturbed jet radius.

\begin{figure}
    \includegraphics[]{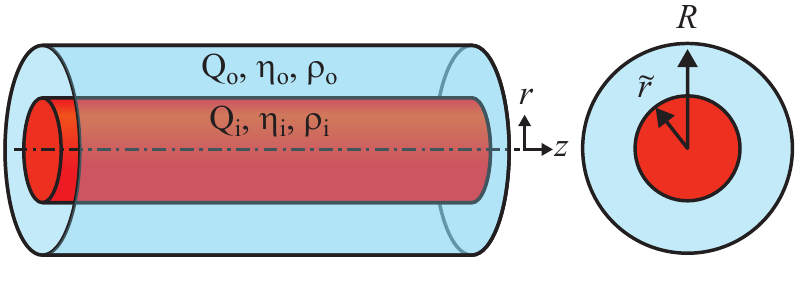}
    \caption{\label{Fig:CH4:coordinates}Coordinate system for an axisymmetric unperturbed inner jet with radius ${\tilde{r}}$ surrounded by a co-flowing liquid in a tube of radius $R$. Subscripts $\mathrm{i}$ and $\mathrm{o}$ refer to the inner and outer phase respectively.}
\end{figure}
In this section we give an expression for the radius of the jet ${\tilde{r}}$ as a function of the inner and outer flow rates, $Q_\mathrm{i}$ and $Q_\mathrm{o}$ respectively, the properties of the two fluids, and the outer capillary tube radius $R$. The coordinate system that represents the initial state of the jet is shown in fig.~\ref{Fig:CH4:coordinates}.

The Reynolds number, as a measure of the relative importance of inertial forces to viscous forces, is typically low in microfluidics, hence the fluid flow can be described using the steady-state Stokes equations for low Reynolds number flows,
\begin{equation}
    \label{Eq:StokesEquation}
    0 = -\nabla{p} + \eta \nabla^2{}u,
\end{equation}
with pressure gradient $\nabla{p}$, viscosity $\eta$, and the velocity $u$. Note also that, with independence of the value of the Reynolds number, equation (\ref{Eq:StokesEquation}) is still valid to describe strictly parallel streams, which is the case under consideration here. The velocity fields are obtained under the assumption of a no-slip boundary condition at the outer tube wall
\begin{equation}
    \label{Eq:BC1}
    u_\mathrm{o}(R) = 0,
\end{equation}
continuity of velocity at the liquid-liquid interface
\begin{equation}
    \label{Eq:BC2}
    u_\mathrm{o}({\tilde{r}}) = u_\mathrm{i}({\tilde{r}}),
\end{equation}
and continuity of shear stress
\begin{equation}
    \label{Eq:BC3}
    \eta_\mathrm{o} \left.\frac{\mathrm{d}u_\mathrm{o}}{\mathrm{d}r}\right|_{r = {\tilde{r}}} = \eta_\mathrm{i} \left.\frac{\mathrm{d}u_\mathrm{i}}{\mathrm{d}r}\right|_{r = {\tilde{r}}}.
\end{equation}
The unperturbed flow of the inner jet resembles a perfect cylinder with constant radius $r = {\tilde{r}}$. The capillary pressure, \emph{i.e.}~the pressure difference across the interface, is given by the Young-Laplace equation
\begin{equation}
    p_\mathrm{i} - p_\mathrm{o} = \frac{\gamma}{{\tilde{r}}},
\end{equation}
with $\gamma$ the interfacial tension. Note that, since the radius of the cylinder is constant, the pressure gradient in both the inner and outer fluid are the same, and equal to $\nabla{p}$.

Inserting the boundary conditions of eq.~(\ref{Eq:BC1}), (\ref{Eq:BC2}), and (\ref{Eq:BC3}) into eq.~(\ref{Eq:StokesEquation}) and integrating gives an expression for the inner and outer liquid velocity profile
\begin{equation}
    u_\mathrm{i}(r) = \frac{\nabla{p}}{4 \eta_\mathrm{i}} \left \{ r^2 - \frac{\eta_\mathrm{i}}{\eta_\mathrm{o}} \left \lbrack 1 + \left( \frac{\eta_\mathrm{o}}{\eta_\mathrm{i}} - 1 \right) \frac{{\tilde{r}}^2}{R^2} \right \rbrack R^2 \right \}\, ,
\end{equation}
\begin{equation}
    u_\mathrm{o}(r) = -\frac{\nabla{p}}{4 \eta_\mathrm{o}} \left( R^2 - r^2 \right)\, .
\end{equation}
The velocity for the inner disperse phase and the outer continuous phase as a function of the radial coordinate are schematically shown in fig.~\ref{Fig:CH4:velocityprofiles}a.

\begin{figure}
    \includegraphics[]{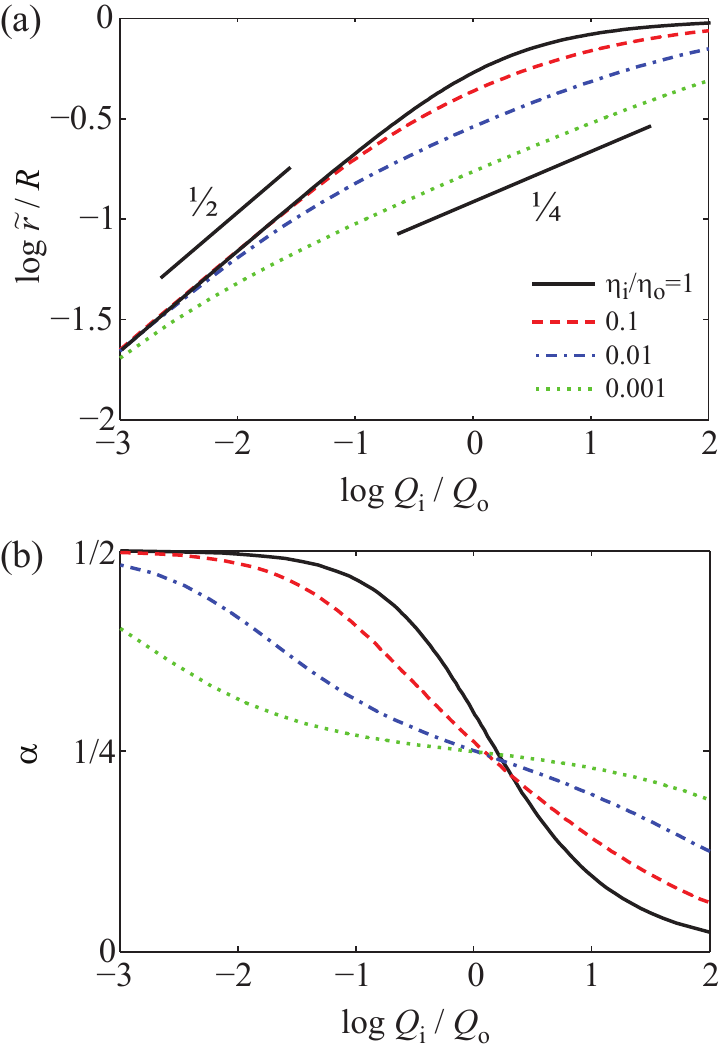}
    \caption{\label{Fig:CH4:roleofviscosity}(a) Normalized radius of the inner jet as a function of the inner and outer volumetric flow rates for various viscosity ratios. The jet's radius decreases, while the maximum inner velocity increases, for decreasing inner-to-outer viscosity ratio. The radius of the jet scales as $\tilde{r} \propto \left( Q_\mathrm{i} / Q_\mathrm{o} \right)^\alpha$.
    (b) Local slope $\alpha = \mathrm{d}{}\log ({{\tilde{r}} / R) } / \mathrm{d}{}\log ({Q_\mathrm{i} / Q_\mathrm{o}})$. In the limit of $Q_\mathrm{i} / Q_\mathrm{o} \rightarrow 0$ the size of the jet scales as $\tilde{r} / R = 2^{-1/2} \left( Q_\mathrm{i} / Q_\mathrm{o} \right)^{1/2}$ independent of the viscosity ratio. For decreasing viscosity ratio ($\eta_\mathrm{i} / \eta_\mathrm{o}$) a scaling $\alpha = 1/4$ becomes pronounced.}
\end{figure}
The volumetric flow rate is found from the integration of the flux over the cross-sectional area of the jet as $Q = 2 \pi{} \int r u(r) \mathrm{d}r$ and, consequently, the flow rate ratio becomes \cite{Utada2005}
\begin{equation}
    \frac{Q_\mathrm{i}}{Q_\mathrm{o}} = \frac{\eta_\mathrm{o}}{\eta_\mathrm{i}} \frac{x^4}{\left( 1 - x^2 \right)^2} + \frac{2 x^2}{1 - x^2},
\end{equation}
with $x = {\tilde{r}} / R$ the dimensionless jet radius.
This equation can of course be inverted, leading to
\begin{equation}
    x = \sqrt{\frac{X}{X + 1}} \mbox{, where } X = \frac{\eta_\mathrm{i}}{\eta_\mathrm{o}} \left( -1 + \sqrt{1 + \frac{\eta_\mathrm{o}}{\eta_\mathrm{i}} \frac{Q_\mathrm{i}}{Q_\mathrm{o}}} \right).
\end{equation}
The dimensionless jet radius $x$ as a function of the flow rate ratio ($Q_\mathrm{i} / Q_\mathrm{o}$) for various inner-to-outer viscosity ratios is depicted in fig.~\ref{Fig:CH4:roleofviscosity}. It is shown here that, for $Q_\mathrm{i} / Q_\mathrm{o} \rightarrow 0$, the size of the jet $\tilde{r} / R = 2^{-1/2} \left( Q_\mathrm{i} / Q_\mathrm{o} \right)^{1/2}$, independently of the viscosity ratio. The size of the jet is thus proportional to the square root of the flow rate ratio, a fact that was reported previously by various groups \cite{Utada2007,Ganan-Calvo2007,Marin2008,Castro2009}.
Note, however, that for the relevant case of bubble formation, \emph{i.e.}~$\eta_\mathrm{i} \ll \eta_\mathrm{o}$, the jet dimensionless radius is given by $\tilde{r} / R \propto (Q_i/Q_o)^\alpha$ with $Q_i/Q_o<1$ and $1/4\leqslant \alpha\leqslant 1/2$ for a large range of values of the flow rate ratio. Indeed, in fig.~\ref{Fig:CH4:roleofviscosity}b the crossover of the local slope $\alpha = \mathrm{d}{}\log ({{\tilde{r}} / R}) / \mathrm{d}{}\log ({Q_\mathrm{i} / Q_\mathrm{o}})$ from the 1/2-scaling to the 1/4-scaling is shown for various viscosity ratios. Only for the limiting case of an inviscid gas, say $\eta_\mathrm{i} / \eta_\mathrm{o} \lesssim 0.001$, does the 1/4 scaling becomes pronounced. Note that a 1/4 scaling exponent is also observed in the co-flow device by {de Castro} \emph{et al.}~\cite{Castro2011}, but under the different conditions of a strongly pressure gradient in the entrance region of the outer capillary.

 For the case of drop formation, the cylindrical liquid jet is unstable against surface perturbations with a wavelength that exceeds the circumference of the jet ($\lambda > 2 \pi {\tilde{r}}^2$) \cite{Chandrasekhar1961}. The fastest growing disturbance $\lambda^\star$ that leads to droplet pinch-off determines the droplet's volume $V = \pi {\tilde{r}}^2 \lambda^\star$, with $\lambda^\star \approx 11.2 {\tilde{r}}$ (for $\eta_\mathrm{i} = \eta_\mathrm{o}$) \cite{Tomotika1935}, hence the size of the droplets $r_\mathrm{d} \approx 2.03 {\tilde{r}} \approx 1.44 R \left( Q_\mathrm{i} / Q_\mathrm{o} \right)^{1/2}$. However, for the case of gas jets ($\eta_\mathrm{i} / \eta_\mathrm{o} \ll 1$), since the period of bubble formation is proportional to $\pi\,R^2 \tilde{r}(Q_i/Q_o)$, bubble volume is calculated through the mass balance $4/3 \pi r^3_b/6\propto \pi\,R^2 Q_i\times \tilde{r}/Q_o$ \cite{Oguz93,PoF07,Castro2011}. This gives $r_b\propto (Q_i/Q_o)^\beta$, with the exponent $\beta$ varying from the two limiting values $\beta=0.5$ for $Q_i/Q_o\rightarrow 0$ ($\alpha=0.5$) and $\beta=5/12$ for $\eta_\mathrm{i} / \eta_\mathrm{o} \rightarrow 0$ ($\alpha=1/4$), being these latter exponents firstly reported by {de Castro} \emph{et al.}~\cite{Castro2011}).

\section{Bubble formation from a `hollow jet'\label{Sec:CH4:gasliquidsystem}}
We now come to the specific case of UCA microbubble formation through the breakup of a `hollow jet'. The co-flowing outer liquid contains high-molecular-weight molecules like surfactants, proteins, and amphiphilic polymers that self-organize at the gas-liquid interface to form a coating around the bubble. This coating, or shell, ensures a much longer bubble lifetime in contrast to uncoated bubbles, therefore improving their stability, an essential condition for medical imaging.

In this section we give an expression for the size of the jet based on the inner gas and the outer liquid flow rates and under the assumption that the interface is rigidified by the surfactant solution \cite{Denkov2009}.

We solve the velocity fields in the Stokes equations (\ref{Eq:StokesEquation}) for an outer liquid flow rate with no-slip condition at the outer tube wall. We assume that the viscous stress imposed by the liquid is entirely balanced by the interface, hence that no stress is transmitted to the gas. This implies a homogeneous pressure distribution inside the jet, hence a flat velocity profile (see fig.~\ref{Fig:CH4:velocityprofiles}b).

The inner and outer velocity fields respectively read
\begin{equation}
    u_\mathrm{i}(r) = \tilde{u} = \mbox{constant, }
\end{equation}
with $\tilde{u}$ the velocity at the interface, and
\begin{equation}
    \label{Eq:CH4:vout}
    u_\mathrm{o}(r) = - \frac{\nabla{p}}{4 \eta_\mathrm{o}} \left(R^2 - r^2\right) + \frac{\log{r/R}}{\log{{\tilde{r}}/R}} \left( \tilde{u} + \frac{R^2 - {\tilde{r}}^2}{4 \eta_\mathrm{o}} \nabla{p} \right).
\end{equation}
Thus, the corresponding volumetric flow rates are given by
\begin{equation}
    Q_\mathrm{i} = \pi {\tilde{r}}^2 \tilde{u}\, \quad \mathrm{and}
\end{equation}
\begin{eqnarray}
    \label{Eq:CH4:flowrate}
    \frac{Q_\mathrm{o}}{2 \pi} &=& \frac{R^4\left(1 - x^2\right)}{16 \eta_\mathrm{o}} \nabla{}p \left( x^2 - 1 + \frac{x^2 - 1 - 2 x^2 \log{x}}{\log{x}} \right) \nonumber\\
    && \qquad + \frac{Q_\mathrm{i}}{\pi x^2} \frac{x^2 - 1 - 2 x^2 \log{x}}{4 \log{x}},
\end{eqnarray}
with $x = {\tilde{r}} / R$.
In this expression the pressure gradient ($\nabla{p}$) and the radius of the jet ($x = {\tilde{r}} / R$) are unknown. Experimentally, both are selected by the system once the gas and liquid flow rates are imposed. Here, it is possible to close the problem by minimizing the viscous energy dissipation $\epsilon$ \cite{Stauffer2005},
\begin{equation}
    \label{Eq:CH4:energydissipation}
    \frac{\epsilon}{2 \pi} = \eta_\mathrm{o} \int_{{\tilde{r}}}^R r \left(\frac{\partial{}u_\mathrm{o}}{\partial{}r}\right)^2 \mathrm{d}{}r \, .
\end{equation}

\begin{figure}
    \includegraphics[]{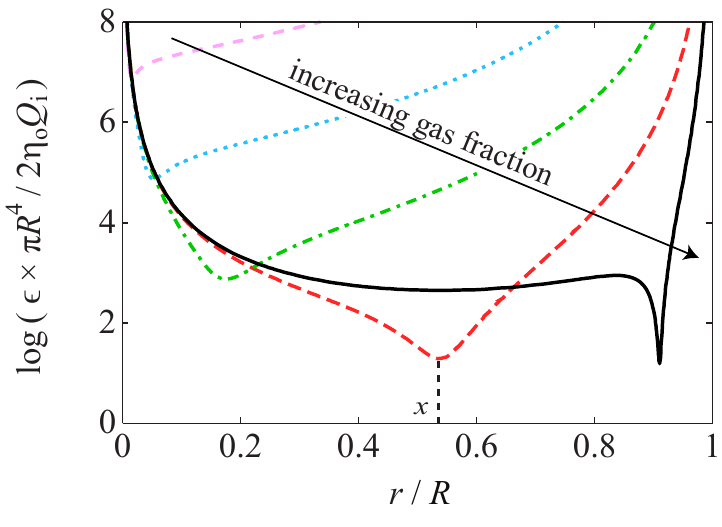}
    \caption{\label{Fig:CH4:energy_dissipation}Energy dissipation $\epsilon$ as a function of the radius of the inner gas jet $r_\mathrm{i}$, for gas-to-liquid flow rate ratios $Q_\mathrm{i} / Q_\mathrm{o}$ equal to 0.001, 0.01, 0.1, 1, and 10 (arrow indicating increasing gas fraction). For each given $Q_\mathrm{i}$ and $Q_\mathrm{o}$ there exists a minimum energy dissipation in the system, which defines the optimum radius of the inner jet $x = {\tilde{r}} / R$.}
\end{figure}
\begin{figure}
    \includegraphics[]{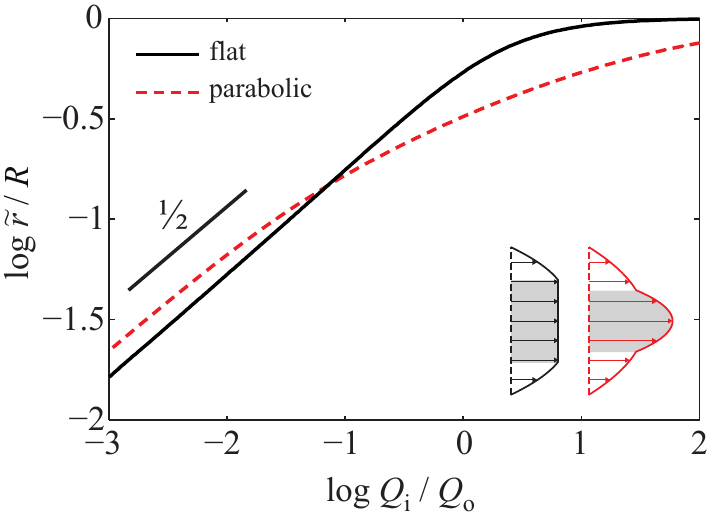}
    \caption{\label{Fig:CH4:comparegasliquid}Optimum radius of the inner gas jet ${\tilde{r}}$ as a function of the gas-to-liquid flow rate ratio $Q_\mathrm{i} / Q_\mathrm{o}$. The solid line represents the optimum radius that is found from minimization of energy dissipation and under the assumption of a flat velocity profile.
    Alternatively, the system is described by two pressure driven co-flowing fluids ($\eta_\mathrm{i}$, $\eta_\mathrm{o}$), with parabolic velocity profiles for both the inner and outer phase for an air-water system ($\eta_\mathrm{o} = 10^{-3}\,\mbox{Pa}\cdot$s, $\eta_\mathrm{i} = 1.8\times{}10^{-5}\,\mbox{Pa}\cdot$s) (dashed line). For $Q_\mathrm{i} / Q_\mathrm{o} \rightarrow 0$, the slope is $1/2$, indicating that the optimum radius scales with the square root of the flow rate ratio (${\tilde{r}} \propto R \left( Q_\mathrm{i} / Q_\mathrm{o} \right)^{1/2}$).}
\end{figure}

Inserting eqs.~(\ref{Eq:CH4:vout}) and (\ref{Eq:CH4:flowrate}) into (\ref{Eq:CH4:energydissipation}) and simplifying gives
\begin{eqnarray}
    \frac{\epsilon}{2 \pi{} \eta_\mathrm{o}} &=& \frac{Q_{\mathrm{i}}^2}{\pi^2 R^4} \left \{ \frac{8 \left( Q_\mathrm{o}/Q_\mathrm{i} - K(x) / {2 x^2}\right)^2}{\left( 1 - x^2 \right) \left( x^2 - 1 + K(x) \right)^2} \right. \nonumber\\
    && \left. \qquad - \frac{1}{x^4 \log{x}} \right \},
\end{eqnarray}
with $K(x) = \left( x^2 - 1 - 2 x^2 \log{x} \right) / \log{x}$.
Thus, the amount of energy that is dissipated into the system depends on both the inner and outer flow rate and the radius of the jet, as is demonstrated in fig.~\ref{Fig:CH4:energy_dissipation}. The curves indicate the dissipation energy for fixed inner and outer volumetric flow rates and its minimum value marks the optimum radius $x$. The optimum radius as a function of the flow rates is found numerically by minimizing the energy dissipation function and is plotted in fig.~\ref{Fig:CH4:comparegasliquid} (solid line).

The gaseous jet is inherently unstable \cite{Guillot2008} and will break up in bubbles by means of either a capillary instability or as a consequence of a pressure drop in the gas. In the case of bubbles are formed as a consequence of a capillary instability, the breakup of the gas jet is driven by the fastest growing disturbance. A linear stability analysis predicts the optimum wavelength for the breakup of a gaseous jet surrounded by a liquid is slightly larger when compared to its inverted system---a liquid jet in air. The stability of a unconfined `hollow jet' in ambient liquid was accomplished by Chandrasekhar \cite{Chandrasekhar1961}, who obtained that the size of the bubbles $r_\mathrm{b} \approx 2.15 {\tilde{r}}\propto (Q_i/Q_o)^{1/2}$. In confined geometries as the ones considered here, the prefactor of this relation will be somewhat different \cite{Tomotika1935,Guillot2008}.
In this case the jet breaks as a consequence of a pressure drop in the gas stream and the bubble formation frequency is proportional to $R^2 {\tilde{r}}/Q_\mathrm{o}$ \cite{Oguz93,PoF07}. The bubble size can then be calculated from the mass balance $r^3_b\propto R^2 {\tilde{r}} Q_i/Q_o\propto R^3 (Q_i/Q_o)^{3/2}$ and, thus, in this case, $r_b/R\propto (Q_i/Q_o)^{1/2}$. Therefore, independent of the source of the instability, the bubble size is  proportional to the square root of the flow rate ratio.

\section{Discussion and conclusion}
In conclusion, we have expressed the radius of gas jets formed in axisymmetric co-flowing streams as a function the control parameters, \emph{i.e.}~the flow rates of both the continuous and the dispersed flows and the outer tube radius. The study has been divided in two parts, depending upon whether the gas-liquid interface contains high-molecular-weight surfactants to avoid the rapid dissolution of the gas into the carrier liquid. In the case where these complex molecules are not present it is found that, if the flow rate ratio is $Q_\mathrm{i} / Q_\mathrm{o} \rightarrow 0$, the diameter of the bubbles is proportional to the square root of the flow rate ratio, independent of the viscosity ratio. However, for finite values of the flow rate ratio $Q_\mathrm{i}/Q_\mathrm{o}<1$ and very low values of the viscosity ratio, \emph{i.e.}~$\eta_\mathrm{i}/\eta_\mathrm{o}\ll 1$, the bubble radius scales as $r_\mathrm{b}\propto (Q_\mathrm{i}/Q_\mathrm{o})^\beta$, with $\beta$ varying from the two limiting values (see fig.~\ref{Fig:CH4:roleofviscosity}) $\beta=0.5$ for $Q_\mathrm{i}/Q_\mathrm{o}\rightarrow 0$ and $\beta=5/12$ for $\eta_\mathrm{i} / \eta_\mathrm{o} \rightarrow 0$. This latter exponent was reported by {de Castro} \emph{et al.}~\cite{Castro2011}.

In the case that high-molecular-weight surfactants are added, we have found that the bubble radius is proportional to the square root of the flow rate ratio. Moreover, in fig.~\ref{Fig:CH4:comparegasliquid} a comparison is made between the gas jet radius predicted based on an inner parabolic velocity profile (dashed line) and a flat velocity profile (solid line) for an air-water system (with viscosity of air $\eta_\mathrm{i} = 1.8\,{\times}\,10^{-5}$\,Pa$\cdot$s and water $\eta_\mathrm{o} = 10^{-3}$\,Pa$\cdot$s). It is demonstrated that for small gas fractions $Q_\mathrm{i} \ll Q_\mathrm{o}$ the influence of the inner flow conditions on the jet radius is marginal.

\acknowledgments
This work is financially supported by the MicroNed technology program of the Dutch Ministry of Economic Affairs through its agency SenterNovem under grant \mbox{Bsik-03029.5}.


\begin{thebibliography}{99}

\bibitem{vanderGraaf2005}
S.~{van der Graaf}, C.~G. P.~H. Schro\"{e}n, and R.~M. Boom.
\newblock Preparation of double emulsions by membrane emulsification---a
  review.
\newblock {\em J. Membr. Sci.}, 251:7--15, 2005.

\bibitem{Hettiarachchi2010}
K.~Hettiarachchi and A.~P. Lee.
\newblock Polymer-lipid microbubbles for biosensing and the formation of porous
  structures.
\newblock {\em J. Colloid Interface Sci.}, 344:521--527, 2010.

\bibitem{Klibanov2002}
A.~L. Klibanov.
\newblock Ultrasound contrast agents: Development of the field and current
  status.
\newblock {\em Top. Curr. Chem.}, 222:73--106, 2002.

\bibitem{Hettiarachchi2007}
K.~Hettiarachchi, E.~Talu, M.~L. Longo, P.~A. Dayton, and A.~P. Lee.
\newblock On-chip generation of microbubbles as a practical technology for
  manufacturing contrast agents for ultrasonic imaging.
\newblock {\em Lab Chip}, 7:463--468, 2007.

\bibitem{Gunther2006}
A.~G\"{u}nther and K.~F. Jensen.
\newblock Multiphase microfluidics: from flow characteristics to chemical and
  materials synthesis.
\newblock {\em Lab Chip}, 6(12):1487--1503, 2006.

\bibitem{Christopher2007}
G.~F. Christopher and S.~L. Anna.
\newblock Microfluidic methods for generating continuous droplet streams.
\newblock {\em J. Phys. D: Appl. Phys.}, 40(19):R319--R336, 2007.

\bibitem{Baroud2010}
C.~N. Baroud, F.~Gallaire, and R.~Dangla.
\newblock Dynamics of microfluidic droplets.
\newblock {\em Lab Chip}, 10(16):2032--2045, 2010.

\bibitem{Nazir2010}
A.~Nazir, C.~G. P.~H. Schro\"{e}n, and R.~M. Boom.
\newblock Premix emulsification: A review.
\newblock {\em J. Membr. Sci.}, 362:1--11, 2010.

\bibitem{Thorsen2001}
T.~Thorsen, R.~W. Roberts, F.~H. Arnold, and S.~R. Quake.
\newblock Dynamic pattern formation in a vesicle-generating microfluidic
  device.
\newblock {\em Phys. Rev. Lett.}, 86(18):4163--4166, 2001.

\bibitem{Chen2009}
C.~Chen, Y.~Zhu, P.~W. Leech, and R.~Manasseh.
\newblock Production of monodispersed micron-sized bubbles at high rates in a
  microfluidic device.
\newblock {\em Appl. Phys. Lett.}, 95:144101, 2009.

\bibitem{Ganan-Calvo1998}
A.~M. {Ga\~{n}\'{a}n-Calvo}.
\newblock Generation of steady liquid microthreads and micron-sized
  monodisperse sprays in gas streams.
\newblock {\em Phys. Rev. Lett.}, 80:285--288, 1998.

\bibitem{Ganan-Calvo2001}
A.~M. {Ga\~{n}\'{a}n-Calvo} and J.~M. Gordillo.
\newblock Perfectly monodisperse microbubbling by capillary flow focusing.
\newblock {\em Phys. Rev. Lett.}, 87:274501, 2001.

\bibitem{Anna2003}
S.~L. Anna, N.~Bontoux, and H.~A. Stone.
\newblock Formation of dispersions using ``flow focusing'' in microchannels.
\newblock {\em Appl. Phys. Lett.}, 82(3):364--366, 2003.

\bibitem{Garstecki2005}
P.~Garstecki, H.~A. Stone, and G.~M. Whitesides.
\newblock Mechanism for flow-rate controlled breakup in confined geometries: A
  route to monodisperse emulsions.
\newblock {\em Phys. Rev. Lett.}, 94(16):164501, 2005.

\bibitem{Raven2006}
J.-P. Raven, P.~Marmottant, and F.~Graner.
\newblock Dry microfoams: formation and flow in a confined channel.
\newblock {\em Eur. Phys. J. B}, 51:137--143, 2006.

\bibitem{Utada2007}
A.~S. Utada, A.~{Fernandez-Nieves}, H.~A. Stone, and D.~A. Weitz.
\newblock Dripping to jetting transitions in coflowing liquid streams.
\newblock {\em Phys. Rev. Lett.}, 99:094502, 2007.

\bibitem{Ganan-Calvo2007}
A.~M. {Ga\~{n}\'{a}n-Calvo}, R.~{Gonz\'{a}lez-Prieto}, P.~{Riesco-Chueca},
  M.~A. Herrada, and M.~{Flores-Mosquera}.
\newblock Nature physics.
\newblock {\em Nature Phys.}, 3(10):737--742, 2007.

\bibitem{Marin2008}
A.~G. Mar\'{i}n, F.~Campo-Cort\'{e}s, and J.~M. Gordillo.
\newblock Generation of micron-sized drops and bubbles through viscous coflows.
\newblock {\em Colloids and Surfaces A: Physicoch. Eng. Aspects}, 344:2--7,
  2009.

\bibitem{Castro2009}
E.~{Castro-Hern\'andez}, V.~Gundabala, A.~{Fern\'andez-Nieves}, and J.~M.
  Gordillo.
\newblock Scaling the drop size in coflow experiments.
\newblock {\em New Journal of Physics}, 11(7):075021, 2009.

\bibitem{Guillot2007}
P.~Guillot, A.~Colin, A.~S. Utada, and A.~Ajdari.
\newblock Stability of a jet in confined pressure-driven biphasic flows at low
  {Reynolds} numbers.
\newblock {\em Phys. Rev. Lett.}, 99(10):104502, 2007.

\bibitem{Utada2005}
A.~S. Utada, E.~Lorenceau, D.~R. Link, P.~D. Kaplan, H.~A. Stone, and D.~A.
  Weitz.
\newblock Monodisperse double emulsions generated from a microcapillary device.
\newblock {\em Science}, 308:537--541, 2005.

\bibitem{Stauffer2005}
D.~Stauffer.
\newblock Teaching {Reiser's} extremal principle for hydrodynamics.
\newblock {\em Am. J. Phys.}, 73(3):282--283, 2005.

\bibitem{Rayleigh1879}
{Lord Rayleigh}.
\newblock On the capillary phenomena of jets.
\newblock {\em Proc. R. Soc. London}, 29:71--97, 1879.

\bibitem{Castro2011}
E.~{Castro-Hern\'andez}, W.~{van Hoeve}, D.~Lohse, and J.~M. Gordillo.
\newblock Microbubble generation in a co-flow device operated in a new regime.
\newblock (under review).

\bibitem{Chandrasekhar1961}
S.~Chandrasekhar.
\newblock {\em Hydrodynamic and hydromagnetic stability}.
\newblock Oxford Press, 1961.

\bibitem{Tomotika1935}
S.~Tomotika.
\newblock On the instability of a cylindrical thread of a viscous liquid
  surrounded by another viscous fluid.
\newblock {\em Proc. Roy. Soc. London}, 150:322--337, 1935.

\bibitem{Denkov2009}
N.~D. Denkov, S.~Tcholakova, K.~Golemanov, K.~P. Ananthpadmanabhan, and
  A.~Lips.
\newblock The role of surfactant type and bubble surface mobility in foam
  rheology.
\newblock {\em Soft Matter}, 5(18):3389--3408, 2009.

\bibitem{Guillot2008}
P.~Guillot, A.~Colin, and A.~Ajdari.
\newblock Stability of a jet in confined pressure-driven biphasic flows at low
  {Reynolds} number in various geometries.
\newblock {\em Phys. Rev. E}, 78:016307, 2008.

\end{thebibliography}

\end{document}